\documentclass[aps,twocolumn,prl,superscriptaddress,preprintnumbers,showpacs,tightenlines]{revtex4}
\usepackage{amssymb}
\usepackage{times}
\usepackage{graphicx}
\usepackage{epsfig}
\usepackage{color,soul}
\usepackage{multirow}
\usepackage{threeparttable}
\usepackage{bm}

\begin{document}
\title{Neutron scattering study of commensurate magnetic ordering in single crystal CeSb$_2$}
\author{Ben-Qiong Liu}
\affiliation{Institute of Nuclear Physics and Chemistry, China Academy of Engineering Physics, Mianyang 621900, China}
\affiliation{J\"ulich Center for Neutron Science JCNS, Forschungszentrum J\"ulich GmbH, Outstation at MLZ, Lichtenbergstra$\beta$e 1, 85748 Garching, Germany}
\author{Li-Ming Wang}
\affiliation{Institute of High Energy Physics, Chinese Academy of Sciences, Beijing 100049, China}
\affiliation{Dongguan Neutron Science Center, Dongguan 523803, China}
\affiliation{J\"ulich Center for Neutron Science JCNS and Peter Gr\"unberg Institut PGI, JARA-FIT, Forschungszentrum J\"ulich GmbH, 52425 J\"ulich, Germany}
\author{Igor Radelytskyi}
\affiliation{J\"ulich Center for Neutron Science JCNS, Forschungszentrum J\"ulich GmbH, Outstation at MLZ, Lichtenbergstra$\beta$e 1, 85748 Garching, Germany}
\author{Yun Zhang}
\affiliation{Science and Technology on Surface Physics and Chemistry Laboratory, Mianyang 621907, China}
\author{Martin Meven}
\affiliation{J\"ulich Center for Neutron Science JCNS, Forschungszentrum J\"ulich GmbH, Outstation at MLZ, Lichtenbergstra$\beta$e 1, 85748 Garching, Germany}
\affiliation{Institute of Crystallography, RWTH Aachen University, Aachen, Germany}
\author{Hao Deng}
\affiliation{Institute of Crystallography, RWTH Aachen University, Aachen, Germany}
\author{Fengfeng Zhu}
\affiliation{J\"ulich Center for Neutron Science JCNS, Forschungszentrum J\"ulich GmbH, Outstation at MLZ, Lichtenbergstra$\beta$e 1, 85748 Garching, Germany}
\author{Yixi Su}
\affiliation{J\"ulich Center for Neutron Science JCNS, Forschungszentrum J\"ulich GmbH, Outstation at MLZ, Lichtenbergstra$\beta$e 1, 85748 Garching, Germany}
\author{Xie-Gang Zhu}
\affiliation{Science and Technology on Surface Physics and Chemistry Laboratory, Mianyang 621907, China}
\author{Shi-Yong Tan}
\affiliation{Science and Technology on Surface Physics and Chemistry Laboratory, Mianyang 621907, China}
\author{Astrid Schneidewind}
\affiliation{J\"ulich Center for Neutron Science JCNS, Forschungszentrum J\"ulich GmbH, Outstation at MLZ, Lichtenbergstra$\beta$e 1, 85748 Garching, Germany}


\begin{abstract}
Temperature and field-dependent magnetization $M(H,T)$ measurements and neutron scattering study of a single crystal CeSb$_2$ are presented. Several anomalies in the magnetization curves have been confirmed at low magnetic field, i.e., 15.6 K, 12 K, and 9.8 K. These three transitions are all metamagnetic transitions (MMT), which shift to lower temperatures as the magnetic field increases. The anomaly at 15.6 K has been suggested as paramagnetic (PM) to ferromagnetic (FM) phase transition. The anomaly located at around 12 K is antiferromagnetic-like transition, and this turning point will clearly split into two when the magnetic field $H\geq0.2$ T. Neutron scattering study reveals that the low temperature ground state of CeSb$_2$ orders antiferromagnetically with commensurate propagation wave vectors $\textbf{k}=(-1,\pm1/6,0)$ and $\textbf{k}=(\pm1/6,-1,0)$, with N\'eel temperature $T_N\sim9.8$ K. This transition is of first-order, as shown in the hysteresis loop observed by the field cooled cooling (FCC) and field cooled warming (FCW) processes.
\end{abstract}

\maketitle

\section{\uppercase\expandafter{\romannumeral1}. INTRODUCTION}


The family of light rare-earth diantimonides $R$Sb$_2$ ($R$=La-Nd) exhibits a remarkably rich variety of behaviors from superconductivity in LaSb$_2$, charge density wave transitions in PrSb$_2$, to anisotropic ferromagnetism in CeSb$_2$. CeSb$_2$ is a well-known example of a complex sequence of several magnetic ordered phases \cite{P.C.Canfield1991,S.L.Budko1998}. It becomes ferromagnetic at about 15.5 K, with the ordered moment oriented within the basal plane, i.e., the easy magnetic direction is perpendicular to the $c$ axis, and then undergoes two or probably three further magnetic transitions at 11.5 K, 9.5 K and 6.5 K at zero field \cite{T.P.Castaneda2013,Y.Zhang2017}. According to transport and magnetization experiments in this compound, there appears to be more than 4 different magnetic phases, some of which are metastable. It is an ideal material for the study of metamagnetism (MMT), which may have a wide variety of technological applications \cite{P.Misra2008} and has attracted much attention following the study of the heavy fermion systems CeRu$_2$Si$_2$ \cite{D.E.MacLauglin1984,J.Flouquet1995}, URu$_2$Si$_2$ \cite{N.Harrison2003}, and CeIr$_3$Si$_2$ \cite{Y.Muro2007}.

Several different interpretations have been proposed for this phenomenon. For examples, Singh \emph{et al.} \cite{R.Singh2013} explained the MMT behavior due to the presence of competing positive and negative exchange interactions. Some research showed that MMT seems to be a crossover between different magnetic states rather than a real transition \cite{R.S.Perry2001}. Another origin could be associated with the abrupt localization of the $f$ electron with a discrete change in Fermi surface volume. In order to determine the electronic properties, Joyce \emph{et al.} \cite{Joyce1992} performed high-resolution synchrotron radiation photoelectron spectroscopy for CeSb$_2$ and found no indication of Kondo-like behavior in the bulk properties. Later, the same research group \cite{Arko1997} showed dispersion of the near-$E_F$ photoemission characteristics which cannot be described by the single-impurity picture \cite{Stewart1984}. On the other hand, the electronic band structure of CeSb$_2$ can be strongly influenced by applying pressure. Kagayama \emph{et al.} \cite{Kagayama2000,Kagayama2005} reported that the MMT field was enhanced by pressure in CeSb$_2$, and showed some features that may be explained as enhancement in the hybridization between the $f$ electron and conduction electrons by decreasing the atomic distance or the lattice parameters.

In spite of considerable effort devoted to the study of MMT behavior, little consensus has been achieved about the mechanism. Just as the crystal structure information is absolutely significant for understanding the physical properties of the crystals, the spin configuration or the magnetic structure information is necessary to understand the magnetic properties. However, to our best knowledge, no detailed magnetic structure as well as the competing interactions in CeSb$_2$ has been reported.

In this work, in addition to the detailed study of the temperature and field-dependent magnetization $M(H,T)$, we performed neutron scattering experiments of CeSb$_2$ to study the magnetic ordering at low temperature. Single crystal neutron diffraction led to the observation of antiferromagnetic peaks described by the propagation vectors $\textbf{k}=(\pm1/6,-1,0)$ and $\textbf{k}=(-1,\pm1/6,0)$, with the N\'eel temperature $T_N\sim$ 9.8 K. Our results confirm the previous prediction \cite{P.C.Canfield1991} that antiferromagnetic exchange is important in CeSb$_2$, and this study might be essential for the understanding of the MMT behaviour.

\section{\uppercase\expandafter{\romannumeral2}. Experiment and analysis}

According to previous structure studies \cite{R.Wang1967,A.Borsese1980}, CeSb$_2$ crystalizes in orthorhombic structure (space group $Cmce$, No.64), with the experimental and calculated lattice parameters listed in Table I. Density functional theory (DFT) calculations were carried out by using the Vienna \emph{ab initio} simulation package (VASP) \cite{Kresse1993,Kresse1996},
with the frozen-core projector augmented wave (PAW) method. We used GGA descriptions for the exchange-correlation functional, with the cutoff energy of 500 meV in a plane-wave basis expansion. The $k$-point meshes in the Brillouin zone (BZ) are sampled by 13$\times$13$\times$4, determined according to the Monkhorst-Pack scheme. The equilibrium volume ($V_0$) is determined by a least-squares fit of the total energy-volume ($E$-$V$) curves to the third-order Birch-Murnaghan equation of state \cite{Birch1947},
\begin{eqnarray}
E(V)&=&E_0+\frac{9V_0B_0}{16}\Big\{\Big[(V_0/V)^{\frac{2}{3}}-1\Big]^3B'_0 \nonumber\\
&+&\Big[(V_0/V)^{\frac{2}{3}}-1\Big]^2\Big[6-4(V_0/V)^{\frac{2}{3}}\Big]\Big\},
\end{eqnarray}
with $E_0$ being the equilibrium energy and $B'_0$ the pressure derivative of $B_0$.

In this work, both nonmagnetic (NM) and ferromagnetic configurations were tested. The total energy for the FM configuration is indeed lower than that of the NM configuration, i.e., in the ground-state a FM state is favored. With the FM configuration, the magnetic moment obtained is about 0.6 $\mu_B$/Ce. 
Compared with the previous X-ray experimental results \cite{R.Wang1967,A.Borsese1980,R.F.Luccas2015} as well as this neutron scattering study, the calculated lattice parameters $a$, $b$, $c$, and the $c/a$ axial ratio show good agreements. 

\begin{table*}[htbp]
\begin{threeparttable}
\caption{Experimental and optimized lattice parameters ($a$, $b$, $c$ in {\AA}), the $c/a$ ratio, and the equilibrium volume ($V_0$, in {\AA}$^3$) of CeSb$_2$. The magnetic calculation was performed for ferromagnetic alignment only.}
\begin{tabular}{llllll}
\toprule
                                       & $a$ ({\AA}) &  $b$ ({\AA}) & $c$ ({\AA})  & $c/a$     & $V_0$ ({\AA}$^3$) \\
\hline
X-ray at 300 K \cite{R.Wang1967}       & 6.295(6)    &  6.124(6)    & 18.21(2)     & 2.893     & 703.20 \\
X-ray at 300 K \cite{A.Borsese1980}    & 6.28-6.30   &  6.13-6.15   & 18.24        & 2.895-2.9 & 705.60 \\
X-ray at 300 K \cite{R.F.Luccas2015}   & 6.28709(15) &  6.16676(13) & 18.2425(3)   & 2.902     & 707.28 \\
X-ray at 80 K \cite{R.F.Luccas2015}    & 6.27043(16) &  6.15017(13) & 18.2126(3)   & 2.905     & 702.35 \\
Neutron at 2.5 K                       & 6.21(4)     &  6.17(1)     & 17.991(1)    & 2.897     & 689.03 \\
DFT-NM                                 & 6.276       &  6.188       & 17.76        & 2.83      & 689.70 \\
DFT-FM                                 & 6.296       &  6.2         & 17.765       & 2.82      & 693.82 \\
\toprule\\
\end{tabular}
\end{threeparttable}
\end{table*}

\subsection{A. Magnetization measurement}
High quality single crystal CeSb$_2$ was grown using the self-flux method \cite{Y.Zhang2017}. The crystals grow as soft plates, with $c$ axis perpendicular to the plate. Magnetization measurements were performed using Quantum Design PPMS DynaCool instrument, with the magnetic field applied parallel or perpendicular to $c$ axis of the crystal, respectively.

\begin{figure}
\centering
\includegraphics[width=8cm]{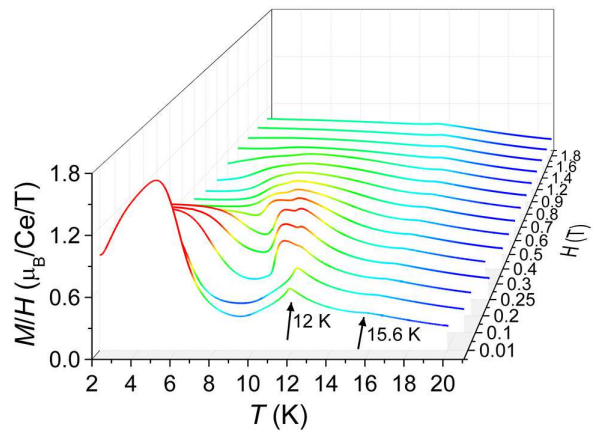}
\hspace{0.5cm}
\caption{(Color online) Magnetic susceptibility of CeSb$_2$ in external magnetic field $H\perp c$ with $H=0.01$, 0.1, 0.2, 0.25, 0.3, 0.4, 0.5, 0.6, 0.8, 0.9, 1.2, 1.4, 1.6, 1.8 T over the temperature range 2 K$-$20 K. These are ZFC data described in the text.}
\end{figure}

Figure 1 shows the magnetic susceptibility versus temperature $T$ (2 K$\leq T\leq$ 20 K) curves measured at different magnetic fields (0.01 T$\leq H\leq$1.8 T) with $H\perp c$. The sample was first cooled in zero-field down to the lowest temperature 2 K and then the magnetic field was applied. These data are denoted as zero-field cooled (ZFC) data. The low temperature behavior of CeSb$_2$ is quite complicated with several phase transitions clearly visible in the magnetization curves. At $H$=0.01 T, a suggested paramagnetic to ferromagnetic phase transition is observed at $T_C=$15.6 K, in good agreement with previous magnetization measurements \cite{P.C.Canfield1991,S.L.Budko1998,Y.Zhang2017}, as well as specific heat curves with a sharp $\lambda$ peak \cite{T.P.Castaneda2013,R.F.Luccas2015}. As the magnetic field increases, this transition will gradually shift to lower temperatures. For examples, at $H=1.8$ T the transition temperature falls to $T_C=$13.1 K.

As the temperature is further reduced, for $H\leq$0.1 T, only one clear broad maximum centred around 12 K is observed as indicated in Fig. 1, which seems to be antiferromagnetic phase transition. It can be reflected in the specific heat curves as a round-shaped peak at $\sim$9.5 K and a small $\lambda$ peak at $\sim$12 K \cite{T.P.Castaneda2013,R.F.Luccas2015}. As the magnetic field increases, this unique broad maximum at around 12 K in the ZFC susceptibility will gradually split into two humps (located at $\sim$9.8 K and $\sim$12 K, respectively) and both of them will shift to lower temperatures. It has been confirmed by neutron scattering that CeSb$_2$ is antiferromagnetic ordering below 9.8 K, which will be described below.


\begin{figure}
\centering
\includegraphics[width=8.5cm]{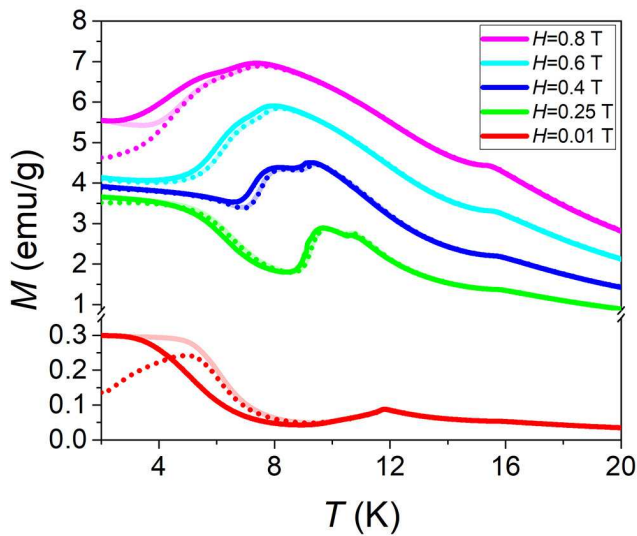}
\hspace{0.5cm}
\caption{(Color online) Magnetization ($M$) versus temperature ($T$) plots for CeSb$_2$ in external magnetic field $H\perp c$ with $H=0.01$, 0.25, 0.4, 0.6, 0.8 T over the temperature range 2 K$-$20 K. The solid lines denote FCC data, the faded lines denote FCW data, and dotted lines denote ZFC data described in the text.}
\end{figure}

Except for these transitions, at low magnetic field $H\leq0.1$ T, the ZFC magnetic susceptibility curve exhibits another hump feature around $T=5$ K, which can be also found in earlier measurements \cite{S.L.Budko1998,Y.Zhang2017}. This feature will be suppressed by increasing magnetic field, and the peak position will slightly shift to a lower temperature. However, for this hump, no corresponding anomaly can be observed in the specific heat measurements \cite{T.P.Castaneda2013,R.F.Luccas2015}. It may suggest the developing of some further complicated spin-spin correlations among Ce ions with decreasing temperature.

Figure 2 shows the magnetization $M$ versus $T$ curves for CeSb$_2$ at selected magnetic fields. The field cooled cooling (FCC) data was collected during cooling down process from high temperature to lowest temperature 2 K in field, and after then the field cooled warming (FCW) data was collected as the temperature increasing again to high temperature. We focus on the interesting features at around 12 K, i.e., single broad peak at low magnetic field $H=0.01$ T but two-peak structure at higher magnetic field $T\geq0.2$ T. It is unambiguously shown in Fig. 2 that there exists a thermal hysteresis between FCC and FCW data, which reveals that this transition is of first order.


The anisotropic susceptibility of CeSb$_2$ is presented in Figure 3, which is in good agreement with the previous results \cite{P.C.Canfield1991,S.L.Budko1998}, with $c$ axis the hard magnetization axis. This remarkable magnetic anisotropy in CeSb$_2$ may be governed by the combination of the crystal field effects \cite{S.L.Budko1998} and the hybridization effect of the 4$f$ wave functions with conduction electrons \cite{K.Shigetoh2007}.

\begin{figure}
\centering
\includegraphics[width=8.5cm]{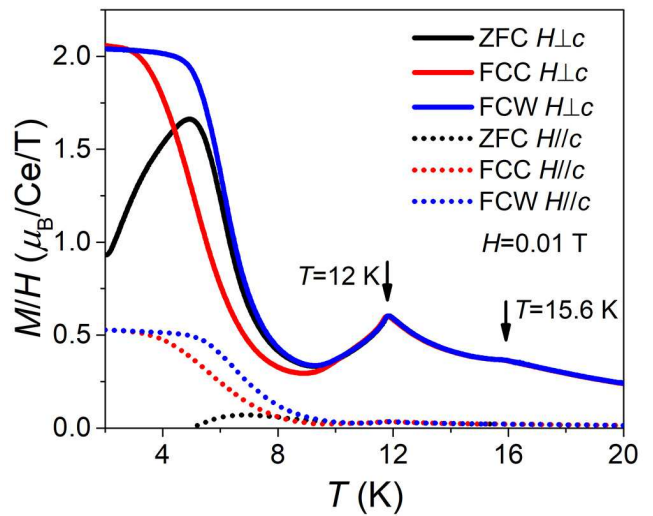}
\hspace{0.5cm}
\caption{(Color online) Magnetic susceptibility of CeSb$_2$ in external magnetic field $H=0.01$ T over the temperature range 2 K$-$20 K, with $H\perp c$ (solid lines) and $H//c$ (dotted lines), respectively.}
\end{figure}

Further insight into the magnetic behavior may be obtained from the isothermal magnetization. Figure 4 shows the magnetization curves as a function of field (0 T$\leq H\leq$4 T) in a temperature range from 2 K to 13 K, where clear metamagnetic transitions are evident for all temperatures. The metamagnetism is broadened as the temperature increases. Each of the $M(H)$ curves was measured with both increasing and decreasing magnetic field. For $T\leq$10 K, $M(H)$ is hysteretic and saturates gradually with increasing $H$ with $M_s$(2 K)$=1.37 \mu_B$/Ce. The step-like changes of magnetization or the plateau behaviors exhibited in the $M(H)$ curves may show different spin-ordering processes like spin flips or rotations occurring with increasing $H$.

At temperature $T=$2 K, several plateaus were observed and at the first plateau $M/M_s\approx0.18$ which is quite close to the value of $1/6$. There are four clear anomalies appearing in the corresponding field derivative of the magnetization d$M$/d$H$ at $H=0.91$, 1.4, 2.35 and 3.5 T.

\begin{figure}
\centering
\includegraphics[width=8.5cm]{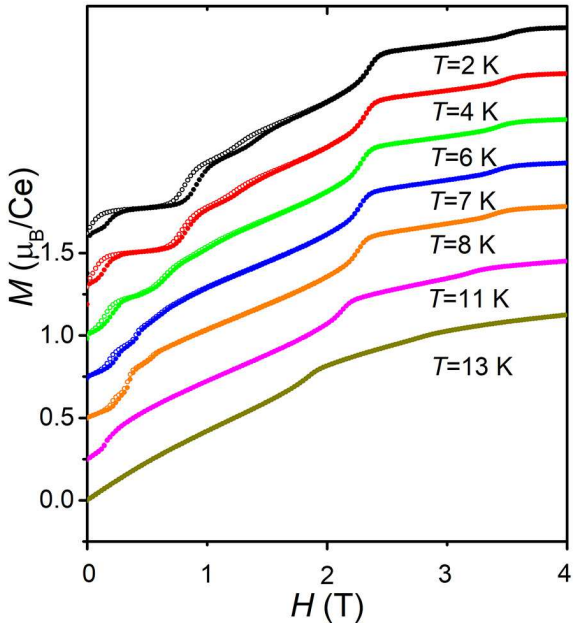}
\hspace{0.5cm}
\caption{(Color online) Isothermal magnetization ($M$) versus magnetic field ($H$) plots for CeSb$_2$ single crystal in temperature range from 2 K to 13 K, with $H\perp c$. 
Based on the curve for $T$=13 K, the other curves for $T=2$, 4, 6, 7, 8, 11 K are shifted above by a step 0.25 between each other for a better view.}
\end{figure}

Based on the measurements of magnetization as a function of temperature $T$ and magnetic field $H$ (Figures $1-4$), a tentative $H$-$T$ phase diagram for CeSb$_2$ for one direction of the magnetic field $H\perp c$ may be constructed, as shown in Figure 5. The phase diagram is really complex with four magnetically ordered phases labeled as I$-$IV, indicating the existence of many competing interactions in CeSb$_2$. As described above that for $H\leq0.1$ T, there is only one broad hump located at about 12 K, which suggests that the phases I and II may be hardly distinguished at low magnetic fields.

\begin{figure}
\centering
\includegraphics[width=8.5cm]{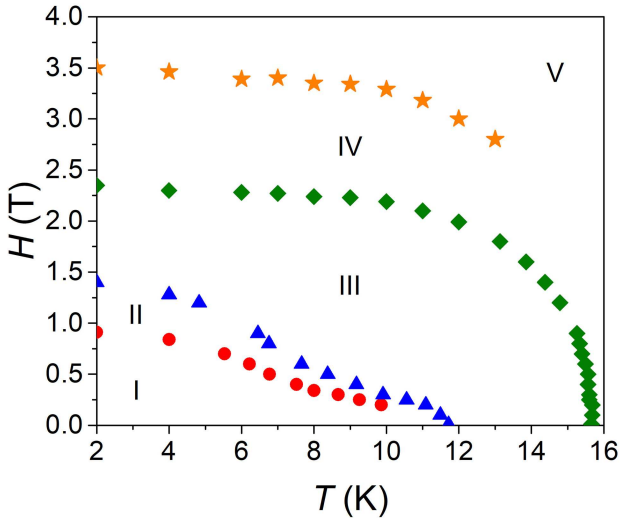}
\hspace{0.5cm}
\caption{(Color online) A proposed $T$-$H$ phase diagram for CeSb$_2$ with $H\perp c$.
}
\end{figure}

\subsection{B. Commensurate magnetic ordering studied by neutron scattering}

In order to understand the magnetic properties in CeSb$_2$, neutron scattering measurements were performed to determine the propagation vector at low temperature. The sample measured is approximately 8$\times$8 mm$^2$ in-plane dimensions and 0.3 mm thickness. Single crystal neutron diffraction experiments were performed on the Diffuse scattering neutron time of flight spectrometer DNS \cite{DNS} and the cold neutron triple-axis spectrometer PANDA \cite{Schneidewind2015} at Heinz Maier-Leibnitz Zentrum (MLZ) in Garching, Germany. Figure 6 shows the difference (3.7 K$-$20 K) neutron diffraction pattern of single crystal CeSb$_2$ measured at DNS with $x$-spin flip mode. Magnetic satellites appearing around the systematically-absent spots $(-1,0,0)$ and $(0,-1,0)$ are clearly shown. The commensurate propagation wave vectors are assumed as $\textbf{k}=(\pm1/6,-1,0)$ and $\textbf{k}=(-1,\pm1/6,0)$.

\begin{figure}
\centering
\includegraphics[width=8.5cm]{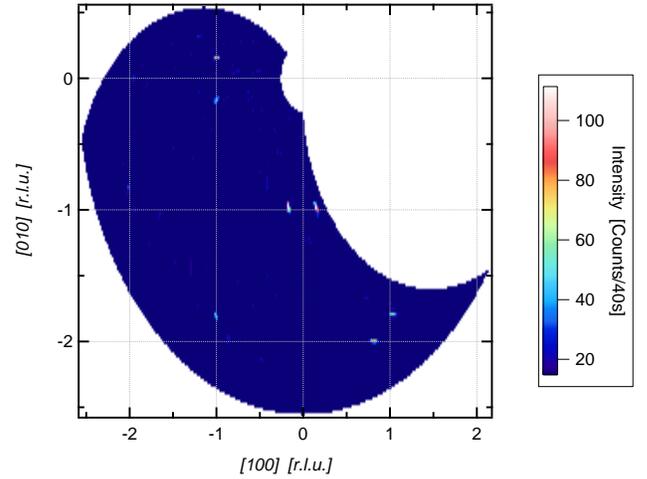}
\hspace{0.5cm}
\caption{(Color online) Difference (3.7 K$-$20 K) neutron diffraction pattern of single crystal CeSb$_2$ measured at DNS with $x$-spin flip mode. Magnetic satellites are observed at ($-1,\pm1/6,0$) and ($\pm1/6,-1,0$). The appearance of a second set of magnetic peaks with $h$ and $k$ flipped can be explained easily by the $Cmce$ structure with $a\approx b$ making it a pseudocubic structure where the large sample size favors the existence of domains rotated by 90$^\circ$ against each other along the common $c$ axis. This has no influence to the temperature dependence of the magnetic peaks studied in this work.}
\end{figure}

Single crystal measurements at PANDA confirmed these features. On Panda, the sample was also aligned in $(h,k,0)$ orientation. Pyrolytic graphite PG(002) was used as monochromator and analyzer, and a Be filter was placed before the analyzer. Elastic scans were performed at fixed neutron wave vectors $k_i=k_f=$1.57 {\AA}$^{-1}$. Figure 7 shows the elastic $h$ scans of the pair of magnetic Bragg peaks ($\pm1/6,1,0$) at $T=5$ K.

\begin{figure}
\centering
\includegraphics[width=8.5cm]{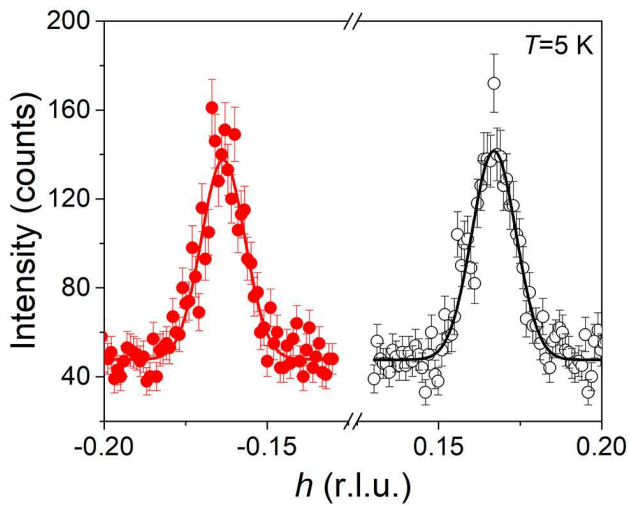}
\hspace{0.5cm}
\caption{(Color online) Elastic $h$ scans of the magnetic satellites ($\pm1/6,1,0$) through the forbidden nuclear reflection $(0,1,0)$ of CeSb$_2$ at $T=5$ K. The solid lines are Gaussian fits.}
\end{figure}

\begin{figure*}
\centering
\includegraphics[width=16cm]{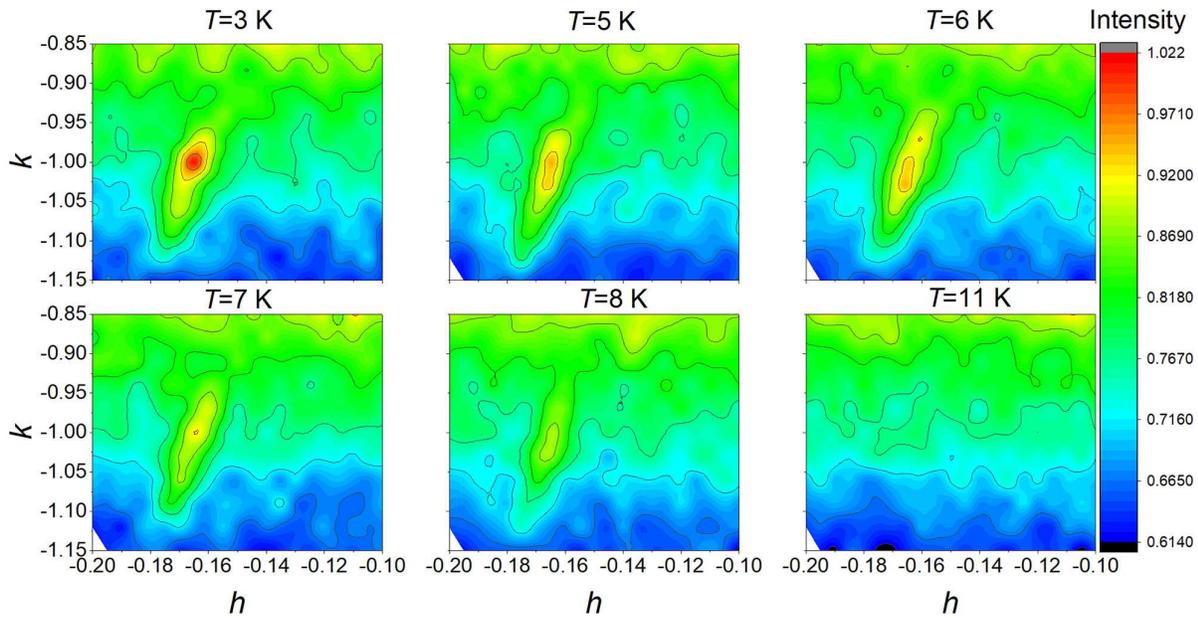}
\hspace{0.5cm}
\caption{(Color online) The map of the $hk0$ plane at different temperatures measured on HEIDI at MLZ, Garching.}
\end{figure*}

A further step in the search for the correct propagation vector was carried out on the single crystal diffractrometer HEIDI at MLZ \cite{HEIDI}. 
As shown in Fig. 8, the mapping of the reciprocal space at different temperatures reveals one clear magnetic reflection which is unambiguously described by the propagation vector $\mathbf{k}=(-1/6,-1,0)$. The propagation vector almost does not change with temperature.

\begin{figure}
\centering
\includegraphics[width=8.5cm]{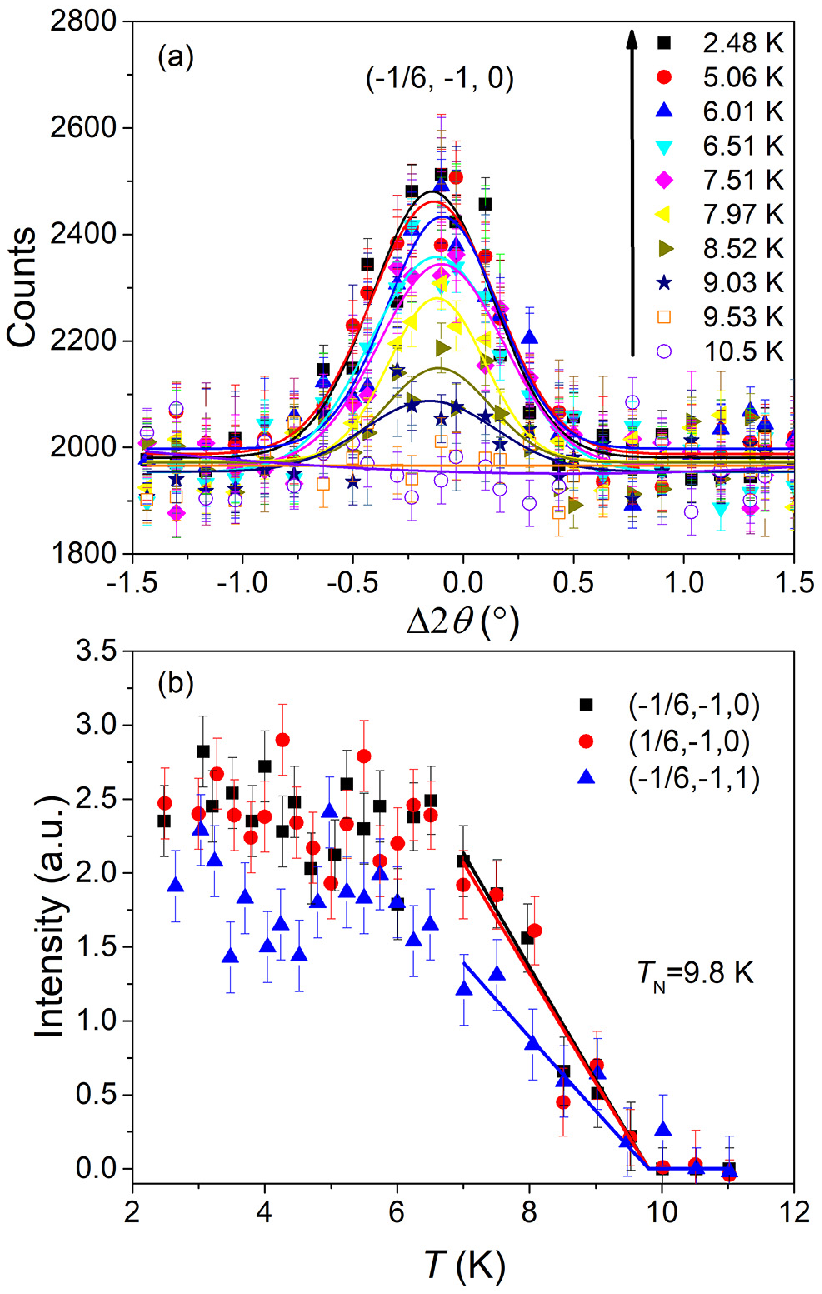}
\hspace{0.5cm}
\caption{(Color online) Temperature dependence of intensity on magnetic reflections ($\pm1/6,-1,0$) and ($-1/6,-1,1$) in CeSb$_2$. The solid lines in (a) are Gaussian fits. The solid lines in (b) are fitted to $I\sim(T_N-T)^{2\beta}$, with $T_N=9.8$ K, and $\beta=0.5$.}
\end{figure}

Figure 9 exhibits the temperature dependence of the intensities for the magnetic reflections ($\pm1/6,-1,0$) and ($-1/6,-1,1$). As the temperature decreased, a magnetic ground state is formed which may be due to the weak hybridization of the $f$ levels with conduction electrons \cite{Arko1997}. The ordered magnetic moment on Ce sites increases with decreasing temperature and remains constant (within the measurement error) at lower temperatures. The fit of the integrated intensity to the power law $I\sim(T_N-T)^{2\beta}$ revealed the ordering temperature $T_N=9.8$ K, corresponding well to $T_N$ obtained from macroscopic magnetization measurements. The fitted value of the critical exponent $\beta=0.5$.

It should be pointed out that except for the phase transitions at around 15.6 K, 12 K, and 9.8 K, the previous resistivity measurements for CeSb$_2$ \cite{Y.Zhang2017} exhibited one more anomaly at 6.5 K which was ascribed to the antiferromagnetic to ferromagnetic transition. However, by neutron scattering no AFM-FM phase transition has been observed down to 2 K. It is possible that the exchange interaction in CeSb$_2$ might be not strong enough to screen out the 4$f$ moments completely at low temperatures, and thus RKKY interaction leads to commensurate antiferromagnetic ordering at $T_N= 9.8$ K.

\section{\uppercase\expandafter{\romannumeral3}. CONCLUSIONS}

CeSb$_2$ displays complex magnetic ordering at low temperature. The temperature $T$ and magnetic field $H$ dependencies of magnetization in CeSb$_2$ exhibit a rich phase diagram which includes four magnetic ordering phases, indicating that there are several tuned exchange interactions. All the kinks associated with the magnetic phase transitions are gradually suppressed to lower temperatures by increasing magnetic field $H$. Moreover, the temperature-dependent magnetization of CeSb$_2$ shows strong anisotropy, which may be due to crystal field effect. Therefore, further inelastic neutron scattering study of CEF excitations may be quite helpful for revealing the hidden mechanism.

Microscopic information on the magnetic structures is essential for understanding the complex magnetic properties of strongly correlated electron systems. A series of neutron scattering measurements were performed on one CeSb$_2$ single crystal. It reveals an antiferromagnetic ground state below $T_N=9.8$ K without any further AFM-FM phase transition down to 2 K. The commensurate antiferromagnetic ordering of CeSb$_2$ can be described by the propagation vectors $\textbf{k}=(-1,\pm1/6,0)$ and $\textbf{k}=(\pm1/6,-1,0)$. Moreover, as known that the Fermi surface topology might play an important role in the determination of the magnetic ordering \cite{Raymond2014}, further DFT calculations as well as experimental technique are required to track the possible modification of Fermi surface.
\section*{ACKNOWLEDGMENTS}

This work is supported by the National Natural Science Foundation of China (Grant No.11875238, 11674406), Science Challenge Project (Grant No.TZ2016004), and Director Foundation of China Academy of Engineering Physics (Grant No.YZ2015009, YZ201501040). B. -Q. Liu has been supported by China Scholarship Council. Part of the data were measured on the single crystal diffractometer HEIDI jointly operated by RWTH Aachen University and JCNS within JARA collaboration. We gratefully acknowledge the computing time on the supercomputer JURECA \cite{Jureca} at Forschungszentrum J\"ulich. We acknowledge helpful discussions with Yuesheng Li and Erxi Feng.



\begin{thebibliography}{99}
\bibitem{P.C.Canfield1991} P. C. Canfield, J. D. Thompson, and Z. Fisk, J. Appl. Phys. \textbf{70}, 5992 (1991).
\bibitem{S.L.Budko1998} S. L. Bud'ko, P. C. Canfield, C. H. Mielke, and A. H. Lacerda, Phys. Rev. B \textbf{57}, 13624 (1998).
\bibitem{T.P.Castaneda2013} T. P\'erez-Casta\~neda, J. Azpeitia, J. Hanko, \emph{et al}., J. Low Temp. Phys. DOI 10.1007/s10909-013-0884-8.
\bibitem{Y.Zhang2017} Y. Zhang, X. Zhu, B. Hu, \emph{et al.}, Chin. Phys. B \textbf{26}, 067102 (2017).
\bibitem{P.Misra2008} P. Misra, \emph{Heay-Fermion Systems}, Elsevier Science, (2007).
\bibitem{D.E.MacLauglin1984} D. E. MacLauglin, C. Tien, W. G. Clark, \emph{et al}., Phys. Rev. Lett. \textbf{53}, 1833 (1984)
\bibitem{J.Flouquet1995} J. Flouquet, S. Kambe, L. P. Regnault, \emph{et al.}, Physica B \textbf{215}, 77 (1995).
\bibitem{N.Harrison2003} N. Harrison, M. Jaime, and J. A. Mydosh, Phys. Rev. Lett. \textbf{90}, 096402 (2003).
\bibitem{Y.Muro2007} Y. Muro, Y. Ohno, T. Okada, and K. Motoya, J. Magn. Magn. Mater. \textbf{310}, 389 (2007)
\bibitem{R.Singh2013} R. Singh, S. K. Srivastava, A. K. Nigam, \emph{et al.}, J. Appl. Phys. \textbf{114}, 243911 (2013).
\bibitem{R.S.Perry2001} R. S. Perry, L. M. Galvin, S. A. Grigera, \emph{et al.}, Phys. Rev. Lett. \textbf{86}, 2661 (2001).
\bibitem{Joyce1992} J. J. Joyce, A. J. Arko, J. Lawrence, P. C. Canfield, Z. Fisk, R. J. Bartlett, and J. D. Thompson, Phys. Rev. Lett. \textbf{68}, 236 (1992).
\bibitem{Arko1997} A. J. Arko, J. J. Joyce, A. B. Andrews, \emph{et al.}, Phys. Rev. B \textbf{56} R7041-R7044 (1997).
\bibitem{Stewart1984} G. R. Stewart, Rev. Mod. Phys. \textbf{56}, 755 (1984).
\bibitem{Kagayama2000} T. Kagayam, G. Oomi, S. L. Bud'ko, P. C. Canfield, Phyica B \textbf{281\&282}, 90-91 (2000).
\bibitem{Kagayama2005} T. Kagayama, Y. Uwatoko, S. L. Bud'ko, P. C. Canfield, Physica B \textbf{359-361}, 320-322 (2005).
\bibitem{R.Wang1967} R. Wang and H. Steinfink, Inorg. Chem., \textbf{6}, 1685 (1967).
\bibitem{A.Borsese1980} A. Borsese, G. Borzone, D. Mazzone and R. Ferro, Journal of the Less-Common Metals, \textbf{79}, 57-63 (1981).
\bibitem{R.F.Luccas2015} R. F. Luccas, A. Fente, J. Hanko, \emph{et al.}, Phys. Rev. B \textbf{92}, 235153 (2015).
\bibitem{Kresse1993} G. Kresse and J. Hafner, Phys. Rev. B: Condens. Matter Mater. Phys., \textbf{47}, 558 (1993)
\bibitem{Kresse1996} G. Kresse and J. Furthm\"uller, Phys. Rev. B: Condens. Matter Mater. Phys., \textbf{54}, 11169 (1996).
\bibitem{Birch1947} F. Birch, Phys. Rev. \textbf{71}, 809 (1947).
\bibitem{K.Shigetoh2007} K. Shigetoh, A. Ishida, Y. Ayabe, \emph{et al.}, Phys. Rev. B \textbf{76}, 184429 (2007).
\bibitem{DNS} Heinz Maier-Leibnitz Zentrum, DNS: Diffuse scattering neutron time-of-flight spectrometer, J. Large-Scale Res. Facil. \textbf{1}, A27 (2015).
\bibitem{Schneidewind2015} A. Schneidewind and P. \v{C}erm\'ak, J. Large-Scale Res. Facil. \textbf{1}, A12 (2015).
\bibitem{HEIDI} Heinz Maier-Leibnitz Zentrum, HEiDi: Single crystal diffractometer at hot source, J. Large-scale Res. Facil. \textbf{1}, A7 (2015). http://dx.doi.org/10.17815/jlsrf-1-20.
\bibitem{Raymond2014} S. Raymond, J. Buhot, E. Ressouche, F. Bourdarot, G. Knebel, and G. Lapertot, Phys. Rev. B \textbf{90}, 014423 (2014).
\bibitem{Jureca} J\"ulich Supercomputing Centre, JURECA: general-purpose supercomputer at J\"ulich supercomputing Centre, J. Large-Scale Res. Facil. \textbf{2}, A62 (2016).


\end{thebibliography}
\end{document}